\documentclass[12pt]{article}

\usepackage{cite}
\usepackage{times}
\usepackage{graphicx} 

\title{A pre-Caloris synchronous rotation for {M}ercury} 

\author
{Mark A. Wieczorek,$^{1\ast}$ Alexandre C. M. Correia,$^{2}$ Mathieu Le Feuvre,$^{3}$ \\
Jacques Laskar,$^{4}$ Nicolas Rambaux$^{5}$\\
\\
\normalsize{$^{1}$Institut de Physique du Globe de Paris, Univ Paris Diderot}\\
\normalsize{4 avenue de Neptune, 94100 Saint-Maur des Foss\'{e}s, France}\\
\normalsize{$^{2}$Departamento de F\'isica da Universidade de Aveiro, Campus Universit\'ario de Santiago}\\
\normalsize{3810-193 Aveiro, Portugal}\\
\normalsize{$^{3}$Laboratoire de Plan\'etologie et G\'eodynamique, Universit\'e de Nantes, France}\\
\normalsize{2 rue de la Houssini\`ere, BP 92208, 44322 Nantes Cedex 3, France}\\
\normalsize{$^{4}$Astronomie et Syst\`{e}mes Dynamiques, {IMCCE-CNRS UMR8028}}\\
\normalsize{Observatoire de Paris, 77 avenue Denfert-Rochereau, 75014 Paris, France}\\
\normalsize{$^{5}$Universit\'e Pierre et Marie Curie--Paris 6; IMCCE-CNRS UMR8028 }\\
\normalsize{Observatoire de Paris, 77 avenue Denfert-Rochereau, 75014 Paris, France}\\
\\
\normalsize{$^\ast$To whom correspondence should be addressed; E-mail: wieczor@ipgp.fr.}
}

\date{}

\begin{document} 

\baselineskip 16pt

\maketitle

{\bf \noindent
The planet Mercury is locked in a spin-orbit resonance where it rotates three times about its spin axis for every two orbits about the Sun \cite{Pettengill+65,Colombo65}. The current explanation for this unique state assumes that the initial rotation of this planet was prograde and rapid, and that tidal torques decelerated the planetary spin to this resonance \cite{Colombo+66, Goldreich+66,Counselman+70,Correia+2004}. When core-mantle boundary friction is accounted for, capture into the 3/2 resonance occurs with a 26\% probability, but the most probable outcome is capture into one of the higher-order resonances \cite{Correia+2009}. Here we show that if the initial rotation of Mercury were retrograde, this planet would be captured into synchronous rotation with a 68\% probability. Strong spatial variations of the impact cratering rate would have existed at this time, and these are shown to be consistent with the distribution of pre-Calorian impact basins observed by Mariner 10 and {\sc Messenger}. Escape from this highly stable resonance is made possible by the momentum imparted by large basin-forming impact events \cite{Melosh75,Lissauer85,Wieczorek+2009}, and capture into the 3/2 resonance occurs subsequently under favourable conditions.
}

Following the discovery that Mercury was in a 3/2 spin-orbit resonance \cite{Pettengill+65,Colombo65}, a number of studies have attempted to explain how this planet might have come to occupy such a particular state. It was shown that as the spin rate of this planet evolved from an initial rapid prograde rotation, capture could occur not only in the 3/2 resonance, but other resonances as well \cite{Colombo+66,Goldreich+66,Counselman+70}. With a molten core \cite{Margot+2007}, capture into many of the higher-order resonances, most notably the 2/1, would occur with high probability \cite{Counselman+70,Peale+77}. Since non-synchronous resonant spin states can be destabilized at times of low orbital eccentricity  \cite{Correia+2004}, and since Mercury's eccentricity is known to vary substantially over time \cite{Laskar2008}, ultimate capture into the 3/2 resonance is made possible. The 26\% probability of ending in the current configuration \cite{Correia+2004,Correia+2009} makes this scenario plausible, but at the same time, somewhat unsatisfactory.

Planetary accretion models imply that the initial spin of the terrestrial planets could have been either prograde or retrograde with equal probability \cite{Dones+93,Lissauer+2000,Kokubo+2007}, leading us to investigate how an initial retrograde rotation of Mercury would have evolved. Given that the orbital evolution of this planet is chaotic, such that it is not possible to predict precisely its evolution beyond a few tens of millions of years, we have performed a statistical study to characterize the likely outcomes of initial retrograde rotation. Using a numerical model that considers tidal dissipation, planetary perturbations, and core-mantle boundary friction \cite{Correia+2009}, the orbital and rotational equations of motion of this planet were integrated for 1000 cases over a 100 million year time period, starting with very close initial conditions and an initial $-$10 day rotational period. 

\begin{figure}
\center
\includegraphics[width=0.75\columnwidth]{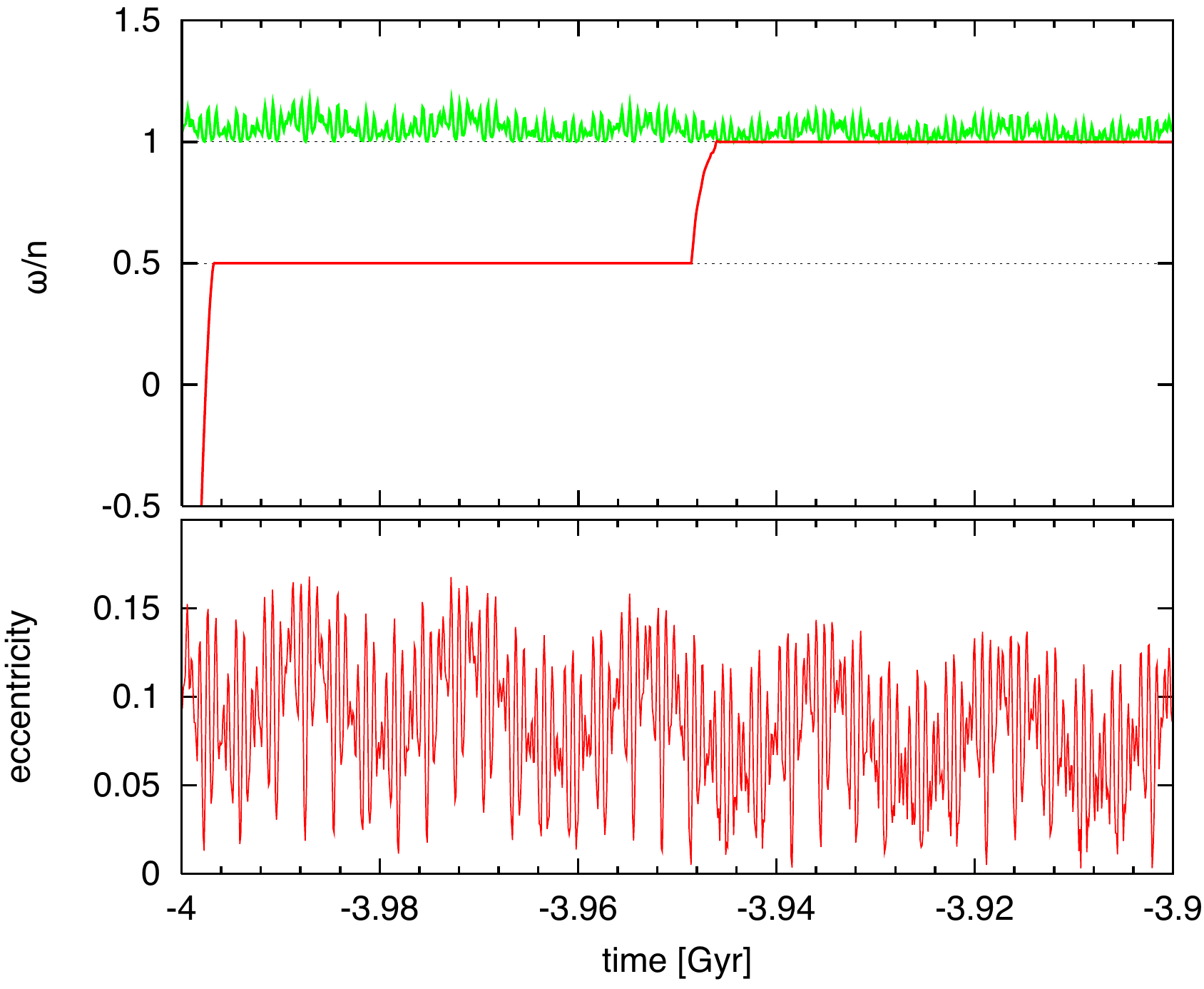}
\caption{Capture into synchronous rotation from initial retrograde rotation. The rotation rate $\omega$ normalized by the orbital mean motion $n$ (top, red), and orbital eccentricity (bottom) are plotted over a 100 million year time period showing capture into the $1/2$ resonance, escape from this resonance at times of low eccentricity, and ultimate capture into synchronous rotation. The equilibrium limit value of the rotation rate (green) is given by a delayed average that damps fast eccentricity-induced variations and is always greater than 1. The statistical results are insensitive to the starting time of the simulations.} \label{fig:capture}
\end{figure}

The theoretical probability of capture in retrograde resonances is extremely low, and no events were observed. The prograde 1/2 resonance is the first important resonance encountered, and capture in this state occurs about 29\% of the time. Some of these captures are destabilized at times of low eccentricity (Fig.~\ref{fig:capture}), and the most likely outcome is to end in synchronous rotation, which occurs with a 68\% probability (Table~\ref{table:capture}). 

Escape from the synchronous resonance cannot occur by variations in eccentricity alone \cite{Correia+2004}, but evolution beyond this state is possible through the momentum imparted to a planet during a basin-forming impact event \cite{Melosh75,Lissauer85,Wieczorek+2009}. Following such an event, the spin rate could have been increased beyond that of the 3/2 resonance, and capture into the present configuration could have occurred as tides decelerated the planet. Alternatively, the impact could have been just large enough to have unlocked the planet from the synchronous resonance, and the spin rate could then have been tidally accelerated to the 3/2 resonance at times of high orbital eccentricity. 

\begin{figure}
\center
\includegraphics{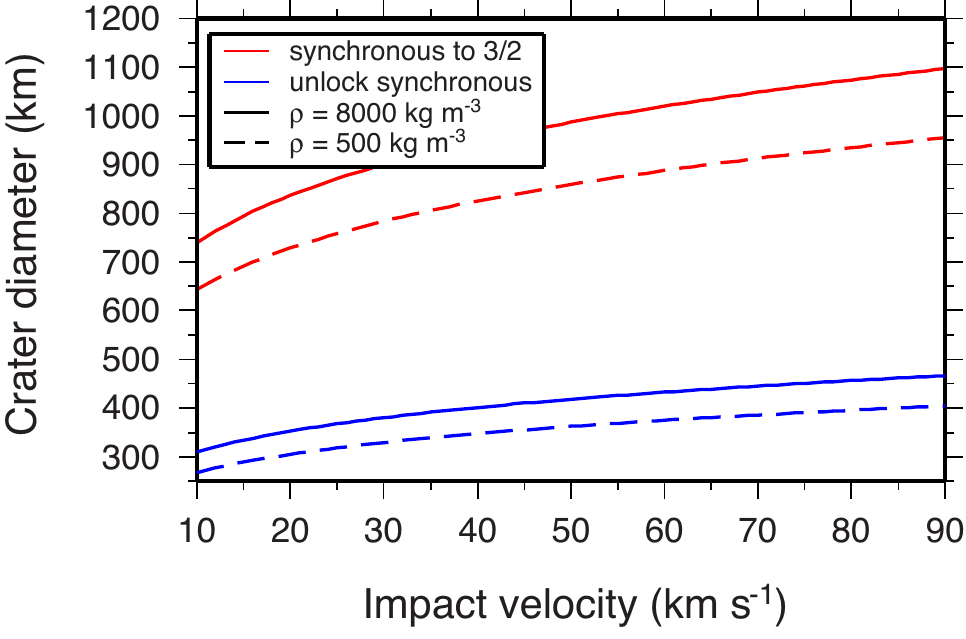}
\caption{Impact crater size and impact velocity required to unlock Mercury from synchronous rotation. Curves in red correspond to the conditions to increased the spin to the 3/2 resonance, and curves in blue represent the conditions necessary to increase the rotation beyond the synchronous resonance. Calculations assume an average impact geometry, and projectile densities of 500 and 8000~kg~m$^{-3}$. The impact velocity on Mercury today ranges from about 10 to 90~km~s$^{-1}$, with an average of 42~km~s$^{-1}$ \cite{LeFeuvre+2011}. The planet is assumed to be completely solid, but if the mantle were uncoupled from the core with no core-mantle friction, the crater diameters would be smaller by about 10\%.
\label{fig:unlock}}
\end{figure}

For a given impact velocity and an average impact geometry, we determine the minimum bolide size for these two unlocking scenarios to occur, and then use standard impact crater scaling laws to estimate the corresponding crater size \cite{Wieczorek+2009}. An event forming a basin with a diameter between about 650 and 1100 km is required for a direct transfer from synchronous rotation to the 3/2 resonance (Fig.~\ref{fig:unlock}). About 14 such basins are known to exist (Supplementary Table~1), but only half of these would have formed with an impact direction that would have increased the rotation rate of the planet, and some of these could have formed before the planet was captured into synchronous rotation. The formation of the 3.73 billion year old Caloris impact basin \cite{LeFeuvre+2011}, which is the largest basin on the planet with a diameter of about 1450 km, would have been the last event capable of performing such a direct transfer. This basin was formed by an oblique impact \cite{Fassett+2009} that likely increased the rotation rate \cite{Wieczorek+2001}, and given its great size, it is possible that this event could have spun the planet up to even higher-order resonances, such as the 4/1. The crater sizes necessary to have just increased the rotation rate of Mercury beyond the synchronous resonance are smaller, between about 250 and 450 km, and about 40 suitably sized basins are known to exist. 

Additional integrations of the rotational and orbital evolution of Mercury were used to quantify the likely outcomes following an impact that destabilizes the synchronous resonance. For three different post-impact rotation rates, 1000 simulations were run starting 3.9 billion years ago and ending at the present day (Supplementary Fig.~1). For the case where the impact just unlocks the planet from synchronous rotation, the planet is recaptured into this resonance 20\% of the time, and capture into the 3/2 resonance occurs 56\% of the time (Table~\ref{table:capture}). When the post-impact rotation rate lies between the 3/2 and 2/1 resonances, capture into the 3/2 resonance is nearly assured with a 96\% probability. When the post-impact rotation rate is greater than that of the 4/1 resonance, the probability of capture into the 3/2 resonance is 26\%, analogous to previous studies that assume initial rapid prograde rotation.

In contrast to previous work, our scenario for capture into the 3/2 resonance makes specific predictions about the density of impact craters on the surface of this planet. Though the cratering rate should be spatially uniform for a planet in non-synchronous rotation, the cratering rates would have been highly asymmetric during periods of synchronous rotation. The trajectories of most asteroids and comets intersect the orbit of Mercury at high angles, and geometric considerations imply that the highest impact rates should occur in the centre of the daylit and unilluminated hemispheres. The Earth also exhibits this behaviour, as is indicated by the radiants of sporadic meteors \cite{CampbellBrown2008}. During synchronous rotation, the axis of the planet's minimum principal moment of inertia would have been directed towards the Sun, and this axis passes through $0^{\circ}$ and $180^{\circ}$ E on the equator \cite{Davies+75}.

\begin{figure}
\begin{center}
\includegraphics[width=0.75\columnwidth]{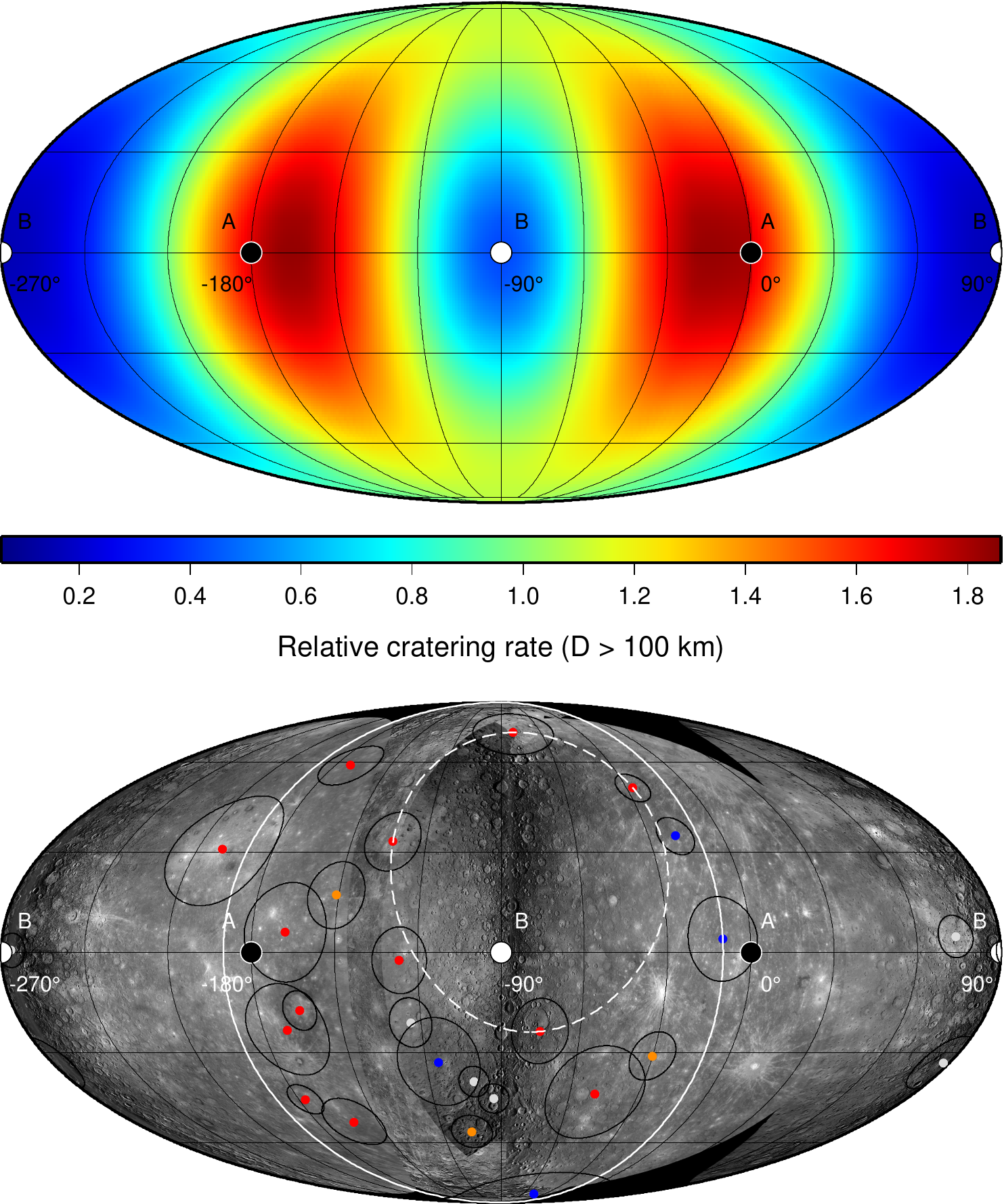}
\end{center}
\caption{Predicted cratering rates and observed impact basins. (top) Synchronous rotation impact cratering rate normalized to the average value for craters larger than 100 km. $0^{\circ}$ longitude corresponds to the subsolar point. (bottom) Impact basins on Mercury with diameters larger than 400 km. Solid red circles (class 1) correspond to the centres of definite basins equal in age or older than the Caloris basin, solid orange circles (class 2) correspond to probable basins \cite{Spudis+88}, solid blue circles (class 3) correspond to possible basins \cite{Spudis+88}, and gray circles (class 4) correspond to basins with uncertain ages with respect to Caloris. None of the centres of these impact basins lie within the dashed white circle. The solid white great circle encompasses the Mariner-10 imaged hemisphere of Mercury. Solid black and white circles mark the axes of the minimum $A$ and intermediate $B$ moments of inertia of Mercury, respectively. Data are plotted on a USGS mosaic of {\sc Messenger} and Mariner 10 images, and is displayed in a Mollweide projection with a central meridian of $90^{\circ}$ W longitude.\label{fig:asymmetry}}
\end{figure}

The expected cratering rates for synchronous rotation are calculated numerically \cite{LeFeuvre+2008,LeFeuvre+2011}. Impact probabilities with a model population of planet-crossing objects \cite{Bottke+2002} were first determined, the relative approach velocities and inclinations were used to determine the impact coordinates on the planet, and scaling laws were used to convert the projectile diameters into crater diameters. Synchronous rotation predicts that the average cratering rate should have varied systematically across the surface by more than a factor of ten (Fig.~\ref{fig:asymmetry}), with the lowest cratering rates being located on the equator at $\pm90^{\circ}$ longitude. The impact velocities vary by a factor of two between the western and eastern hemispheres, and similar results are obtained using the orbital elements of the known near-Earth asteroids (Supplementary Fig.~2).

We test whether spatial variations in the ancient impact-cratering rate ever existed on Mercury by analyzing the distribution of large impact basins, the vast majority of which are older than the Caloris basin. Impact basins identified by Mariner-10 based geologic mapping \cite{Schaber+77,Frey+79,Spudis+88} and {\sc Messenger} flyby images \cite{Fassett+2011} (Supplementary Table~1) are utilized. Since the spatial density of craters smaller than 100 km in diameter have been affected by more recent geologic processes \cite{Fassett+2011}, and since the numerous craters that are somewhat larger than this might be in saturation \cite{Fassett+2011}, we use only those basins that are larger than 400 km (Fig.~\ref{fig:asymmetry}). These are divided into four classes based on the reliability of their identification and stratigraphic age. The first three contain only those basins that are known to be equal in age or older than Caloris, whereas the last corresponds to basins that have an unknown or uncertain relative stratigraphic age with respect to this event. 

The distribution of large impact basins is striking. First, there are very few large basins on the hemisphere from 0 to 180$^{\circ}$ E longitude. This is consistent with the variations predicted by synchronous rotation, but as the solar illumination conditions over much of the flyby images of this hemisphere are sub-optimal for crater identification \cite{Fassett+2011}, this agreement may be equivocal. Second, there is a clear deficit of basins centred near the equator and 90$^{\circ}$ W, which is also consistent with synchronous rotation. No basin centres are located within a circle with a $48^{\circ}$ angular radius close to this point, even though the low sun elevations of the {\sc Messenger} images in this region are favourable to the detection of impact basins.

The significance of these observations is quantified using Monte Carlo simulations of the average angular distance of basins to the point (90$^{\circ}$ W, 0$^{\circ}$ N). Only those basins that lie on the Mariner-10 imaged hemisphere are analyzed since this hemisphere has been subjected to detailed geologic mapping, and since this hemisphere has been imaged under favourable illumination conditions \cite{Fassett+2011}. The probability that the observed value would occur by chance is only 3\% when considering all possible pre-Calorian basins, and increases to 6\% when using the smaller number of most reliably identified basins. The inclusion of basins with uncertain stratigraphic ages increases the probability to 11\%, but this is to be expected as some portion of these likely formed after the Caloris impact when this planet was certainly in non-synchronous rotation. The observed value of the average angular distance to (90$^{\circ}$ W, 0$^{\circ}$ N) is nearly identical to that predicted for synchronous rotation, and similar results are obtained using basins with diameters greater than 300 km (Supplementary Tables~2-3 and Supplementary Fig.~3). Craters with diameters less than 300 km show no such asymmetry, consistent with crater saturation \cite{Fassett+2011}.

If Mercury was ever in a state of synchronous rotation, one hemisphere would have been extremely cold, and the other extremely hot. Substantial quantities of volatile deposits would have accumulated on the unilluminated hemisphere, just as is believed to have occurred within the permanently shadowed craters near the poles \cite{Harmon+2001}. If these deposits were thick enough, it might be possible to find relict geomorphological differences between the two ancient hemispheres, as well as hydrous minerals that might have formed when these deposits were melted. The large differences in temperature between the two hemispheres would have given rise to spatial variations in the thickness of the lithosphere, and viscous relaxation of surface features would have occurred more quickly on the daylit hemisphere. Since the average impact velocity would have differed between the two hemispheres by a factor of about two, the volume of impact melt generated during an impact would have varied by a factor of about five \cite{Pierazzo+1997}. The consequences of synchronous rotation will be illuminated by data obtained from NASA's mission {\sc Messenger} \cite{Solomon2003}.

\clearpage

\clearpage

\begin{table}
 \caption{Probability of capture in spin-orbit resonances. Probabilities are given in percent for initial retrograde rotation and three different post-impact spin rates. $\omega_0$ is the initial angular rotation rate and $n$ is the orbital mean motion.
\label{table:capture}}    
\begin{center}  
\begin{tabular}{| c | c | c c c|} \hline
spin-orbit   & \multicolumn{1}{c|}{pre impact} & \multicolumn{3}{c|}{post impact}  \\  
resonance & $\omega_0/n = -8.8$ & $\omega_0/n = 1.1 $ & $ \omega_0/n = 1.8  $ & $ \omega_0/n = 4.4  $ \\
\hline
5/1 &  0.1 &    $-$ &	$-$ &	$-$   \\ 
9/2 &  0.1 &    $-$ &	$-$ &	$-$   \\
4/1 &  $-$ &    $-$ &	$-$ &	$-$   \\
7/2 &  $-$ &    $-$ &   $-$ &	4.7   \\
3/1 &  $-$ &    0.1 &   0.3 &  11.6   \\ 
5/2 &  0.1 &    0.2 &   0.1 &  22.1   \\
2/1 &  0.4 &    0.7 &   2.7 &  31.6   \\
3/2 &  2.3 &   56.2 &  96.0 &  25.9   \\
1/1 & 68.1 &   20.4 &   0.4 &   3.9   \\
1/2 & 28.9 &    $-$ &   $-$ &   $-$   \\
none&  $-$ &   22.4 &   0.5 &   0.2   \\ 
\hline
\end{tabular}
\end{center}
\end{table}

\clearpage
\mbox{}
\vskip 2in
\begin{center}
\section*{Supplementary Information} 
\end{center}

\clearpage

\begin{table*}
{\center Supplementary Table 1. Impact basins on Mercury with diameters greater than 300 km.\\}
{\footnotesize
\label{table:data}
\begin{tabular}{lcccccl}
\hline
Name & Latitude$^{a}$ & Longitude$^{a}$ & Diameter$^{a}$ & Age$^{b}$ & Class$^{c}$ & Source\\
& $^{\circ}$N &  $^{\circ}$E & km \\
\hline
Caloris 					& 30.9 	& 159.7 	& 1456.7 	& C 	& 1 & \cite{Schaber+77,Frey+79,Spudis+88,Fassett+2011} \\
Andal-Coleridge 			& -43 	& -49 	& 1300 	& pT	& 1 & \cite{Spudis+88}\\
Tir 						& 6 		&  -168 	& 1250 	& pT & 1 & \cite{Spudis+88}\\
Eitoku-Milton 				& -23 	& -171 	& 1180 	& pT & 1 &  \cite{Spudis+88}\\
Bartok-Ives 				& -33 	& -115 	& 1175 	& pT & 3 & \cite{Spudis+88}\\
Donne-Moliere 				& 4 		& -10. 	& 1060 	& pT & 3 &  \cite{Spudis+88}\\
Sadi-Scopas 				& -83 	& -44 	& 930 	& pT	& 3 & \cite{Spudis+88}\\
Matisse-Repin				& -24 	& -75 	& 850 	& pT & 1 & \cite{Spudis+88}\\
Budh 					& 17 	& -151 	& 850 	& pT	& 2 & \cite{Frey+79,Spudis+88}\\
Sobkou 					& 33.4	& -133.5 	& 785.3 	& pT	& 1 & \cite{Schaber+77,Frey+79,Spudis+88,Fassett+2011}\\
Borealis 					& 72.1 	& -80.9 	& 785.2 	& pT	& 1 &  \cite{Spudis+88,Fassett+2011}\\
Mena-Theophanes 			& -1 		& -129 	& 770 	& pT	& 1 & \cite{Schaber+77,Spudis+88}\\
Rembrandt 				& -33.1 	& 87.7 	& 696.7 	& $\sim$C & 4 & \cite{Fassett+2011,Watters+2009}\\
Vincente-Yakovlev 			& -52.6 	& -162.1 	& 692.5	& pT & 1 & \cite{Spudis+88,Fassett+2011}\\
Ibsen-Petrarch 				& -31 	& -30 	& 640	& pT	& 2 & \cite{Spudis+88}\\
Beethoven 				& -20.8 	& -123.9 	& 632.5 	&  $\sim$C &  4 & \cite{Schaber+77,Frey+79,Spudis+88,Fassett+2011}\\
Brahams-Zola 				& 59 	& -172 	& 620	& pT & 1 & \cite{Spudis+88}\\
Tolstoj 					& -17.1 	& -164.6 	& 500.6 	&	T & 1 & \cite{Schaber+77,Frey+79,Spudis+88,Fassett+2011}\\
Hawthorne-Riemenschneider	& -56 	&  -105 	& 500	& pT	& 2 & 	\cite{Spudis+88}\\
Gluck-Holbein 				& 35 	& -19 	& 500 	& 	pT	& 3 & \cite{Spudis+88}\\
Dostoevskij 				& -45.0 	& -176.2	& 430.4 	&  pT  & 1 & \cite{PlanetNomenclature,Schaber+77,Frey+79,Spudis+88}\\
(unnamed) 				& 0.6 	& 93.4 	& 428.4 	& & 4 & \cite{Fassett+2011}\\
(unnamed) 				& -39.0 	& -101.4 	& 420.3 	& & 4 & \cite{Fassett+2011}\\
(unnamed) 				& -44.5 	& -93.2 	& 411.4 	& & 4 & \cite{Fassett+2011}\\
Derzhavin-Sor Juan 			& 50.8 	& -26.9 	& 406.3 	&	pT	& 1 & \cite{Schaber+77,Spudis+88,Fassett+2011}\\
(unnamed) 				& -2.6 	& -56.1 	& 392.6 	& & 4 &  \cite{Fassett+2011}\\
(unnamed) 				& 27.9 	& -158.6 	& 389.0 	& pC & 1 & \cite{Schaber+77,Frey+79,Fassett+2011,Guest+83}\\
Vyasa 					& 50.7 	&  -85.1 	& 379.9 	&	pC	& 1 &  \cite{Fassett+2011,McGill+83}\\
Shakespeare 				& 48.9 	& -152.3	& 357.2 	& 	pT	& 1 & \cite{Schaber+77,Frey+79,Spudis+88,Fassett+2011}\\
Hiroshige-Mahler 			& -17.0 	& -23.0 	& 340.3 	&	pT	& 1 & \cite{Frey+79,Spudis+88,Fassett+2011}\\
Chong-Gauguin 			& 57.1 	& -107.9 	& 325.6 	&	pT	& 1 & \cite{Spudis+88,Fassett+2011}\\
Raphael 					& -20.3 	&  -76.1 	& 320.4 	& pC & 1 &  \cite{Schaber+77,Frey+79,Fassett+2011,King+90}\\
Goethe 					& 81.5 	& -54.3 	& 319.0 	&	pC	& 1 & \cite{Schaber+77,Frey+79,Fassett+2011,Grolier+84}\\
(unnamed) 				& -2.5 	& -44.6 	& 311.4 	& & 4 &  \cite{Schaber+77,Frey+79,Fassett+2011}\\
(unnamed) 				& 28.9 	& -113.8 	& 307.9 	& & 4 &  \cite{Schaber+77,Fassett+2011}\\
(unnamed) 				& -25.0 	& -98.8 	& 307.6 	& & 4 & \cite{Fassett+2011}\\
(unnamed) 				& -17.3 	& -96.8 	& 303.4 	& & 4 &  \cite{Fassett+2011}\\
\hline
\end{tabular}

{\footnotesize
$^{a}$ Locations and diameters are from the most recent source. One impact basin tabulated by \cite{Schaber+77} (27.3$^{\circ}$ N, 146.1$^{\circ}$ E, D=379 km), but not identified in other studies, is not included. Two basins, Homer (-36.5$^{\circ}$ E, -1.4$^{\circ}$ N) and an unnamed crater (101.78$^{\circ}$ E,  70.23$^{\circ}$ N), that have measured diameters less than 300 km by \cite{Fassett+2011}, but larger than this by \cite{Baker+2011}, are not included.

$^{b}$ C, Calorian; T, Tolstojan; pT, pre-Tolstojan. pC refers to craters that are presumed to be older than the Caloris basin based on mapped c1 or c2 degradation states \cite{McCauley+81}.

$^{c}$ Class 1 corresponds to definite basins of \cite{Spudis+88} and craters from \cite{Fassett+2011}, class 2 corresponds to probable basins from \cite{Spudis+88} and class 3 corresponds to possible basis from \cite{Spudis+88}. All class 1--3 basins are younger than, or equal in age to, the Caloris basin, whereas class 4 basins have uncertain or undefined stratigraphic ages with respect to Caloris.}
}
\end{table*}

\clearpage

\begin{table*}
Supplementary Table 2. Observed average angular distance $\bar{\theta}$ of basins on the Mariner-10 imaged hemisphere from the point ($90^{\circ}$ W, $0^{\circ}$ N), and the probability (in percent) of obtaining this value by chance from a uniform cratering rate. Craters with diameters less than $300$ km taken from \cite{Fassett+2011}.\\
\begin{center}
\begin{tabular}{cccccc}
\hline
Diameter range, km & Class & Number & $\bar{\theta}$  & $P(>\bar{\theta})$ \\
\hline
$>400$ & $1$  & 13 & 67.5 & 6.2  \\
$>400$ & $1-2$ & 16 & 66.3 & 6.7  \\
$>400$ & $1-3$  & 20 & 67.1 & 3.1 \\
$>400$ & $1-4$ &25 & 63.6 & 10.9  \\
$>300$ & $1$ & 20 & 64.7 & 7.0 \\
$>300$ & $1-2$ & 23 & 65.2 & 7.0 \\
$>300$ & $1-3$ &  27 & 65.5 & 3.5 \\
$>300$ & $1-4$ & 37 & 58.5 & 42.7 \\
$300-400$ & $1-4$ & 12 & 49.1 & 90.7 \\
$200-300$ &$-$ & 24 & 57.5 & 52.8 \\
$100-200$ &$-$  & 215 & 55.2 & 95.3 \\
$20-100$ & $-$ & 3131& 54.6 & 100\\
\hline
\end{tabular}
\end{center}
\end{table*}

\clearpage

\begin{table*}
Supplementary Table 3.  Expected average angular distance $\bar{\theta}$ of basins on the Mariner-10 imaged hemisphere to the point ($90^{\circ}$ W, $0^{\circ}$ N). Two cases are given for synchronous rotation as the subsolar point could correspond to either 0$^{\circ}$ or 180$^{\circ}$ longitude.\\

\begin{center}
\begin{tabular}{lc}
\hline
Case & $\bar{\theta}$ in degrees \\
\hline
Uniform cratering & 57.7 \\
Synchronous rotation$^a$ (apex = 90$^{\circ}$ W) & 62.6 \\
Synchronous rotation$^a$ (apex = 90$^{\circ}$ E) &  68.9 \\
Synchronous rotation$^b$ (apex = 90$^{\circ}$ W) & 67.0 \\
Synchronous rotation$^b$ (apex = 90$^{\circ}$ E) &  64.2 \\
\hline

\small $^a$ Using the near-Earth asteroid model of \cite{Bottke+2002}.\\

\small $^b$ Using the observed near-Earth asteroids with $H<18$ \cite{astorb}.

\end{tabular}
\end{center}
\end{table*}

\clearpage

\begin{figure}
\begin{center}
\includegraphics[width=0.9\columnwidth]{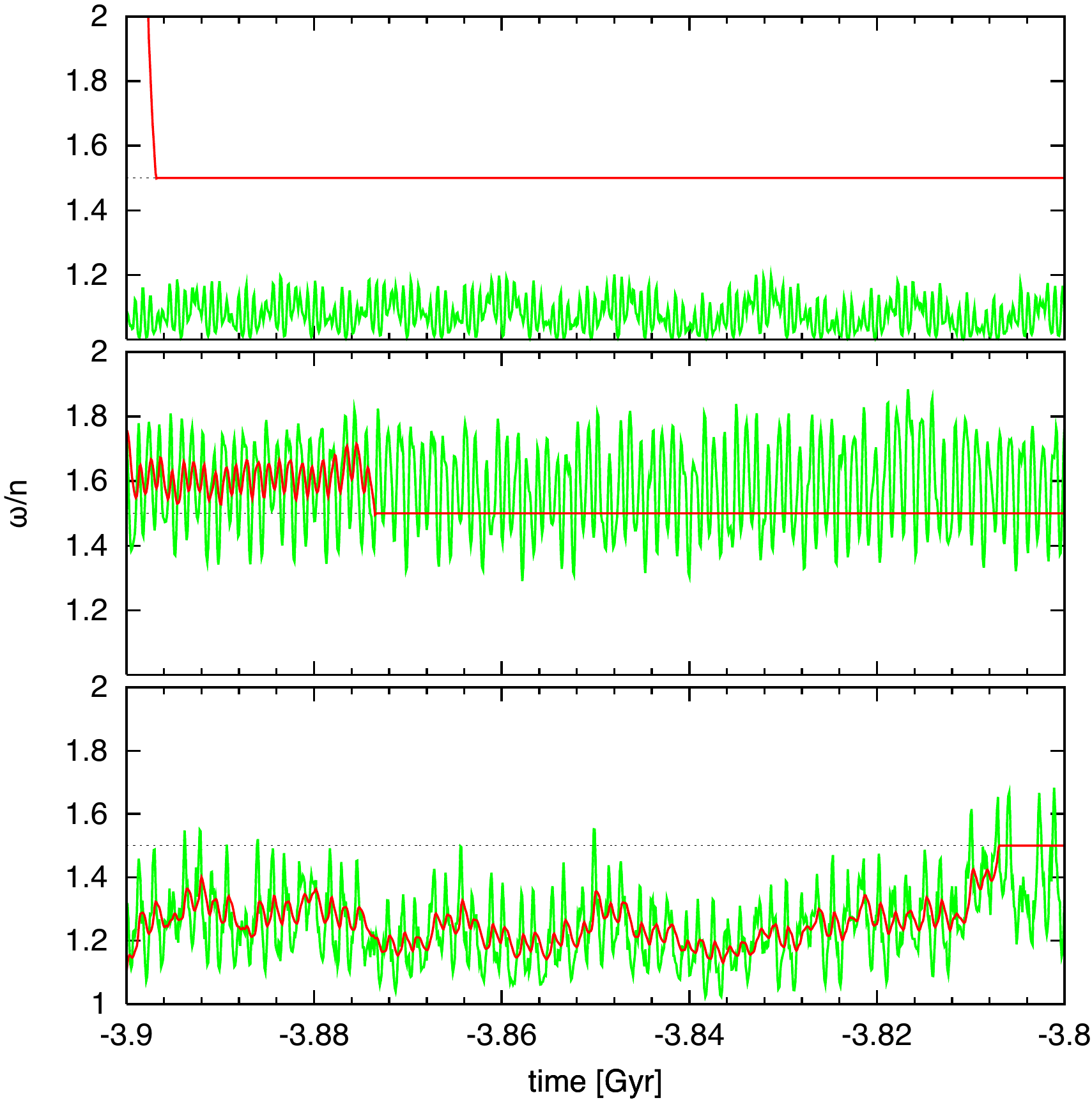}
\end{center}

Supplementary Figure 1. Typical cases of capture into the $3/2$ spin-orbit resonance. The rotation rate (red) and limit value (green) are plotted as a function of time following an impact event. Scenario $(a)$ corresponds to an increase in the rotation rate to $\omega_0/n=4.4$. This is the classical scenario of capture, where the eccentricity is always less than $0.285$ and the limit value of the rotation rate is always lower than that of the $3/2$ resonance. Scenario $(b)$ corresponds to an increase in the rotation rate  to $\omega_0/n=1.8$, just above that of the 3/2 resonance. In this scenario, the rotation rate tracks the limit value until it crosses and is captured into the $3/2$ resonance. Scenario $(c)$ corresponds to the case where the post-impact rotation rate was just large enough to have escaped synchronous rotation, with $\omega_0/n=1.1$. In this case, the rotation rate tracks the limit value until it crosses and is captured into the $3/2$ resonance.
\end{figure}

\clearpage

\begin{figure}
\begin{center}
\includegraphics[width=\columnwidth]{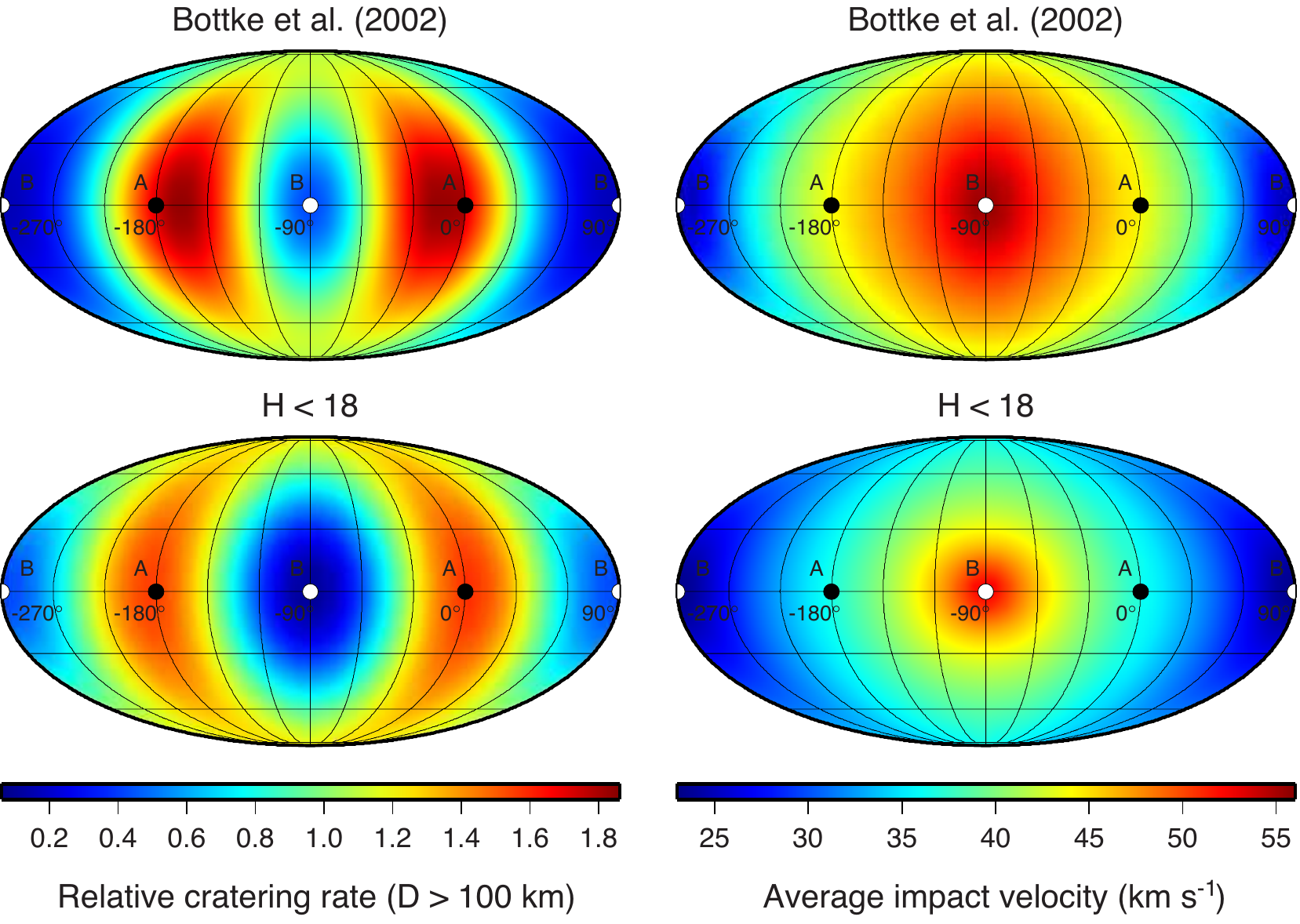}
\end{center}

Supplementary Figure 2. Predicted impact cratering rate normalized to the average value for craters greater than 100 km in diameter (left), and predicted average impact velocity (right) for synchronous rotation. The upper two images were obtained using the Bottke et al. \cite{Bottke+2002} model population of near-Earth objects, and the lower two images were obtained using the orbital elements of the known near-Earth objects with absolute magnitudes less than 18\cite{astorb}. $0^{\circ}$ longitude corresponds to the average subsolar point. Solid black and white circles mark the axes of the minimum $A$ and intermediate $B$ moments of inertia of Mercury, respectively. Cratering rates are only marginally sensitive to the employed crater diameter.
\end{figure}

\clearpage

\begin{figure}
\begin{center}
\includegraphics[width=\columnwidth]{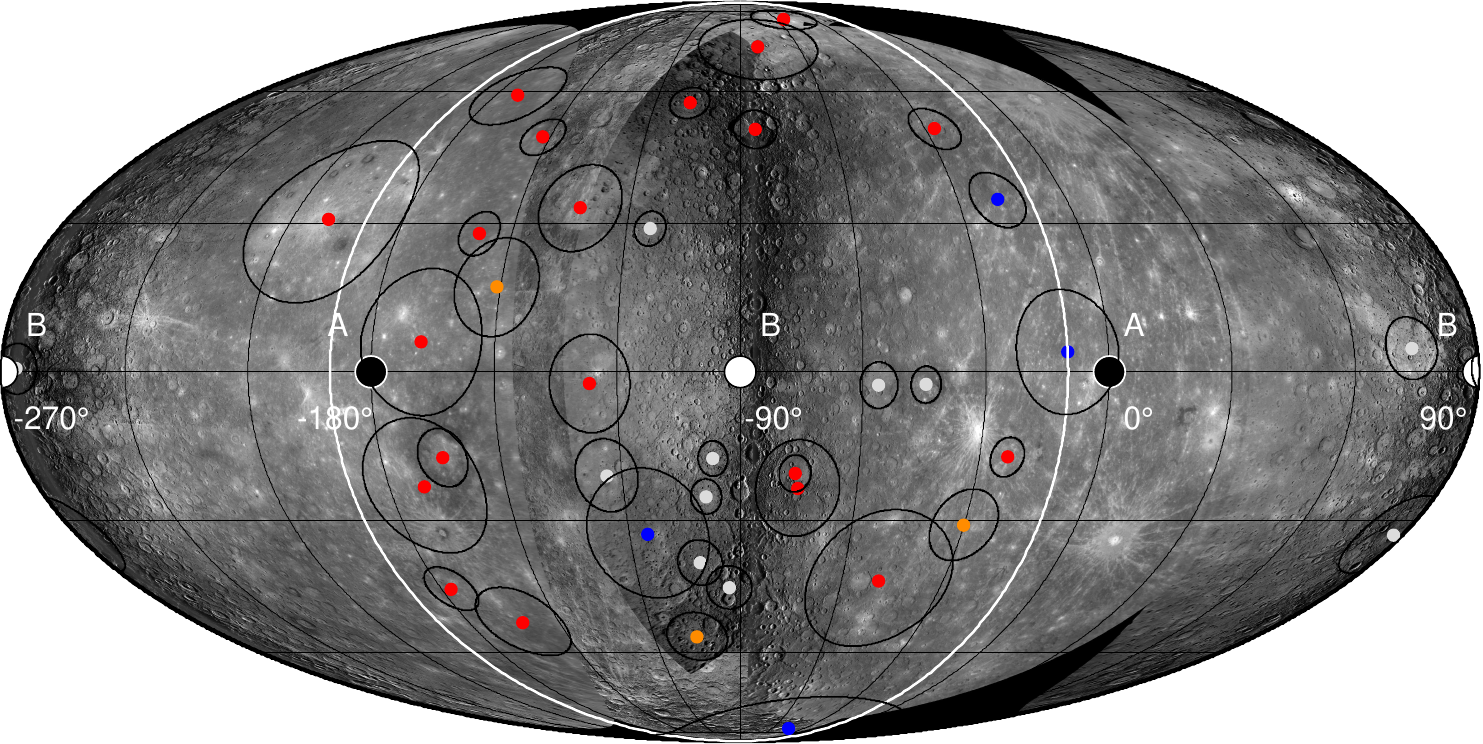}
\end{center}

Supplementary Figure 3. Impact basins on Mercury with diameters larger than 300 km (data from Supplementary Table~1). Solid red circles correspond to the centers of definite basins equal in age or older than the Caloris basin (class 1), solid orange circles correspond to probable basins (class 2), solid blue circles correspond to possible basins (class 3), and gray circles correspond to basins with uncertain ages with respect to Caloris (class 4). The solid white great circle encompasses the Mariner-10 imaged hemisphere of Mercury. Solid black and white circles mark the axes of the minimum $A$ and intermediate $B$ moments of inertia of Mercury, respectively. Data are plotted on a USGS mosaic of {\sc Messenger} and Mariner 10 images, and is displayed in a Mollweide projection with a central meridian of $90^{\circ}$ W longitude.
\end{figure}

\end{document}